\documentclass[a4paper,11pt]{article}
\usepackage{}
\usepackage{mathrsfs}
\usepackage{amsfonts}
\usepackage{amsmath}
\usepackage{amssymb}
\usepackage{indentfirst}
\usepackage{cite}
\usepackage[title]{appendix}
\usepackage{graphicx,url,psfrag}
\usepackage{tikz}
\usetikzlibrary{decorations.pathreplacing,calc,decorations.fractals,through}
\newcommand{\pa}{\partial}

\newcommand{\al}{\langle}
\newcommand{\ar}{\rangle}

\newtheorem{theorem}{Theorem}
\newtheorem{corollary}{Corollary}
\newtheorem{lemma}{Lemma}
\newtheorem{proposition}{Proposition}
\title{Quasi-periodic solutions to the negative-order KdV hierarchy}
\author {Jinbing Chen\thanks{cjb@seu.edu.cn;}\\ 
School of Mathematics, Southeast University,\\
Nanjing, Jiangsu 210096, P.R. China\\
}
\date{}
\begin{document}
\maketitle
\begin{abstract}\noindent
A complete algorithm is developed to deduce quasi-periodic solutions for the negative-order KdV (nKdV) hierarchy by using the backward Neumann systems. From the nonlinearization of Lax pair, the nKdV hierarchy is reduced to a family of backward Neumann systems via separating temporal and spatial variables. The backward Neumann systems are shown to be integrable in the Liouville sense, whose involutive solutions yield the finite parametric solutions of nKdV hierarchy. The negative-order Novikov equation is given, which specifies a finite-dimensional invariant subspace of nKdV flows. By the Abel--Jacobi variable, the nKdV flows are integrated with Abel--Jacobi solutions on the Jacobi variety of a Riemann surface. Finally, the Riemann--Jacobi inversion of Abel--Jacobi solutions is studied, from which some quasi-periodic solutions to the nKdV hierarchy are obtained.
\end{abstract}

\section{Introduction}
\label{sec:1}
The nKdV equation takes the form \cite{1}
\begin{equation}\label{1.1}
u_{t_{-2}}+2v_x=0,\qquad v_{xxx}+4uv_x+2u_xv=0,
\end{equation}
which was originally proposed by Verosky in view of the negative direction of recursion operators for symmetries. It connects with the sinh-Gordon equation through the Miura's transformation \cite{1}, as well as the sine-Gordon equation under the reciprocal transformation \cite{2}. Let $\rho$ be the kernel of Schr\"{o}dinger operator $\partial^2+u$ and $v=\rho^2$, where $\partial=\partial/\partial x$ and $\partial^2=\partial^2/\partial x^2$, etc. The nKdV equation (\ref{1.1}) can be put into its equivalent form \cite{3}
\begin{equation}\label{1.2}
\left(\frac{\rho_{xx}}{\rho}\right)_{t_{-2}}-4\rho\rho_x=0, \quad {\rm or} \quad
\rho_{xxt_{-2}}\rho-\rho_{xx}\rho_{t_{-2}}-4\rho^3\rho_x=0.
\end{equation}
Followed by Verosky's idea, four sets of symmetries together with the nKdV hierarchy were presented associated with the KdV system \cite{3,4}. Subjected to the Painlev\'{e} test, Karasu--Kalkani et al. arrived at an integrable system of KdV6 equation from a class of sixth-order nonlinear wave equations \cite{5}. It is worthy of pointing out that the KdV6 equation is in fact the integrable coupling of KdV and nKdV flows. From the configuration of KdV6, Kupershmidt depicted it as a nonholonomic deformation of bi-Hamiltonian system (i.e. the Kupershmidt deformation) \cite{6}, and Zhou further generalized the nonholonomic perturbation to the mixed hierarchy of soliton equations, which includes the Kupershmidt deformation as its special member \cite{7}. Recently, the Hamiltonian structure, infinitely many conservation laws, $N$-soliton, $N$-kink and
quasi-periodic wave solutions have been derived for the nKdV equation \cite{8,9}. It is seen that soliton and kink solutions occur simultaneously in the nKdV equation. Also, the nKdV equation is found to be linked with the Camassa-Holm (CH) equation via a reciprocal transformation \cite{10}, such that the nKdV equation becomes a hot spot of integrable nonlinear evolution equations (INLEEs). Therefore, this paper is intend to design a constructing scheme for getting quasi-periodic solutions to the nKdV hierarchy characterized by the nKdV equation (\ref{1.1}).

\par
Due to the negative powers of Olver's operator, the original form of nKdV equation is of a nonlocal INLEE, where the nonlocality is eliminated by introducing an additional spectral potential $v$ or $\rho$. As known so far, most attention is paid to positive-order INLEEs (pINLEEs), while the negative-order INLEEs (nINLEEs) are not studied as comprehensively as the positive-order ones. The main reason lies in the fact that the nonlocality of nINLEEs is out of the control of usual methods of analysis. Even in a long time, there was no a fascinating example of nINLEEs of physical interests in the community of mathematical physics. After the physical derivation of CH equation \cite{11}, the contented tranquility of such a subject has been completely disturbed ever since. It is found that the CH equation possesses not only the non-smooth soliton solutions (peakons) \cite{11,12}, but also the presence of wave breaking \cite{13}. Soon afterwards, the interest in the study of non-smooth solitons leads to the advent of peakon systems of Degasperis-Procesi (DP), CH2, and some others with cubic nonlinearity \cite{14,15,17,18}.  Seemingly, the peakon systems are described as local INLEEs by the Helmholtz operator $1-\partial^2$. Indeed, they are all residing in the negative-order hierarchies stemmed from linear spectral problems, which in a sense enrich the mathematical structures and the physical backgrounds of nINLEEs. Apart from the peakon systems themselves belong to the category of nINLEEs, they have reciprocal links to some previously known nINLEEs, for example, the CH and nKdV \cite{10}, the DP and negative-order Kaup-Kupershmidt \cite{2}, the CH2 and negative-order AKNS equations \cite{15}, and so on. Though nINLEEs are intractable to be handled due to the nonlocality, it is the nonlocality that extremely enhances the variety of nINLEEs, such as the presence of peakons, cuspons, and breaking waves etc \cite{11,12,13,15,17,18}. As a result, we focus on the integrable structure, the evolution of phase flows and some other explicit solutions for nINLEEs.

\par
Within the regime of mathematical physics, the KdV equation is almost the most popular and best studied nonlinear model, being integrable at both the classical and the quantum level. Apart from itself interesting, it is intimately related to the scattering theory of one-dimensional Schr\"{o}dinger equation \cite{19}. Moser and Trubowitz used one-dimensional Schr\"{o}dinger equation to demonstrate a connection between the classical Neumann problem and the finite-gap solution of KdV equation \cite{20,21}. Furthermore, the linear operator with a self-consistent potential was introduced by Cherednik to construct finite-gap solutions for the nonlinear Schr\"{o}dinger equation \cite{22}, and developed in the framework of the theory of integrable operators with self-consistent coefficients \cite{23,24}. As ususal, one considers infinite-dimensional integrable systems (IDISs) to be solved in such a way as splitting into several finite-dimensional integrable systems (FDISs) that are easier to be treated with some available tools. Based on the works \cite{19,20,21,22,23,24}, the nonlinearization of Lax pair makes it possible to reduce INLEEs to FDISs of solvable ordinary differential equations \cite{25}, from which soliton solutions, quasi-periodic solutions, and rogue periodic waves can be deduced by separating temporal and spatial variables \cite{26,27,28,29,30,31,32}. Recently, it is found that nINLEEs can be reduced to a family of backward Neumann type systems on the symplectic submanifold \cite{33}. The Neumann map not only generates the finite-gap potential of spectral problem, but also specifies a finite-dimensional invariant subspace for the negative-order flow. Note that the nKdV and positive-order KdV (pKdV) hierarchies share the spatial part of Lax pairs, i.e. the one-dimensional Schr\"{o}dinger equation. It stimulates us to decompose the nKdV hierarchy into backward Neumann systems, and take the backward Neumann systems as a basis to deduce quasi-periodic solutions of the nKdV hierarchy.

\par

The main purpose of this paper is to generalize the application of backward FDISs for getting explicit solutions to nonlocal INLEEs. The mechanism of our treatment involved in can be outlined into threefold. The first step is the finite-dimensional integrable reduction, such that the nKdV hierarchy is decomposed into a class of backward Neumann system on the tangent bundle of unit sphere. The relationship between the nKdV hierarchy and the backward Neumann systems is established, where involutive solutions of the backward Neumann systems yield finite parametric solutions of the nKdV hierarchy. Besides, the negative-order Novikov equation is proposed, which specifies a finite-dimensional invariant subspace for the nKdV flows. The restriction of nKdV flows onto the solution space of negative-order Novikov equation become the backward Neumann flows. The next step is to straighten out the nKdV flows on the Jacobi variety of a Riemann surface. The starting point in this procedure is the introduction of a set of elliptic variables for the backward Neumann systems, whose dynamics are controlled by the Dubrovin-type equations. And, by an Abel map from the divisor group to the Jacobi variety, the Able--Jacobi (angle) variable is elaborated, such that the nKdV flows are integrated with Abel--Jacobi solutions represented by the linear combinations of flow variables. The third step is to explore the Riemann--Jacobi inversion of Abel--Jacobi solutions. It is noted that the spectral potential can be expressed as the symmetric functions of elliptic variables. From the asymptotic expansions of normalized
holomorphic differentials and Riemann theta functions, the Abel--Jacobi solutions of nKdV flows are transformed to the original spectral potential in terms of the Riemann theorem. Actually, we take the nKdV hierarchy as an example to illustrate the construction of quasi-periodic solutions through backward Neumann systems, from which an effective way is presented to solve nonlocal INLEEs in view of backward FDISs.

\par
This paper is organized as follows. In section \ref{sec:2}, with an infinite sequence of Lenard gradients, we unify the pKdV and nKdV hierarchies into a generating pattern. In section \ref{sec:3}, the finite-dimensional integrable reduction of nKdV hierarchy is studied on the tangent bundle of unit sphere. The Liouville integrability of backward Neumann systems is completed in section \ref{sec:4}. Section \ref{sec:5} is to connect the nKdV hierarchy with the backward Neumann systems, and section \ref{sec:6} is devoted to the algebro-geometric construction of explicit solutions for the nKdV hierarchy. Finally, in the appendix A, some new results associated with the Kupershmidt deformation of KdV hierarchy are presented depending upon the current work.


 \setcounter{equation}{0}
\section{The bidirectional KdV hierarchy}
\label{sec:2}
It often happens that starting from a properly chosen spectral problem, soliton equations emerge with an infinite sequence of higher-order members \cite{34}, and can be formally represented by the Lenard gradients together with the Lenard operator pair $K$ and $J$. If taking the kernel of $J$ as initial values, we arrive at a hierarchy of pINLEEs; on the other hand, defining the kernel of $K$ as initial values, we come to a hierarchy of nINLEEs. Since all the pINLEEs and nINLEEs share the spatial part of Lax representations and a pair of bi-Hamiltonian operators, the union of pINLEEs and nINLEEs is thus identified as a bidirectional hierarchy. To make the paper self-contained, let us first retrieve the bidirectional KdV (bKdV) hierarchy in our setting, and further supply some formulae for later use.

\par
Our point of departure is the one-dimensional Schr\"{o}dinger equation that coincides with the $2\times2$ matrix form as
\begin{equation}
\varphi_x=U\varphi, \quad U=\sigma_2-(\lambda+u)\sigma_3, \quad
\varphi=\left(
\varphi_1,\varphi_2\right)^T,\label{2.1}
\end{equation}
with
\begin{equation}\sigma_1=\left(\begin{array}{cc}
1&0\\
0&-1\\
\end{array}\right),\quad
\sigma_2=\left(\begin{array}{cc}
0&1\\
0&0\\
\end{array}\right),\quad
\sigma_3=\left(\begin{array}{cc}
0&0\\
1&0\\
\end{array}\right),\label{2.2}
\end{equation}
where $\lambda$ is a spectral parameter, and $u$ is a spectral potential. We solve the stationary zero-curvature equation of spectral problem (\ref{2.1})
\begin{equation}
V_x=[U,V],\quad  V=\left\{\begin{array}{l}
a\sigma_1+
b\sigma_2+c\sigma_3=\sum\limits_{j\geq 0}(a_j\sigma_1+
b_j\sigma_2+c_j\sigma_3)\lambda^{-j},\\
-a\sigma_1-
b\sigma_2-c\sigma_3=-\sum\limits_{j\leq -1}(a_j\sigma_1+
b_j\sigma_2+c_j\sigma_3)\lambda^{-j},\\
\end{array}
 \right.\label{2.3}
\end{equation}
and then derive
\begin{equation}\label{2.4}
a_{jx}=b_{j+1}+ub_j+c_j,\quad b_{jx}=-2a_j,\quad c_{jx}=-2a_{j+1}-2ua_j,\quad j\in\mathbb{Z},
\end{equation}
\begin{equation}\label{2.5}
4\partial b_{j+1}=-(\partial^3+2(\partial u+u\partial))b_j.
\end{equation}
Let $a_0=b_0=0$ and $c_0=-1$ be the initial seeds. By the recurrence chains (\ref{2.4}) and (\ref{2.5}), together with the kernel of Schr\"{o}dinger operator $\rho$, we derive a few terms of $a_j$, $b_j$, and $c_j$
\begin{equation}
\label{2.6}\begin{array}{c}
a_{-1}=-\rho\rho_x,\qquad b_{-1}=\rho^2,\qquad c_{-1}=-\rho_x^2,\\
a_{-2}=2\partial\rho^2\partial^{-1}\rho^{-2}\partial^{-1}\rho^{-2}\partial^{-1}\rho^{2}\partial\rho^{2},
\quad b_{-2}=-4\rho^2\partial^{-1}\rho^{-2}\partial^{-1}\rho^{-2}\partial^{-1}\rho^{2}\partial\rho^{2},\\
c_{-2}=2(\partial^2+2u)\rho^2\partial^{-1}\rho^{-2}\partial^{-1}\rho^{-2}\partial^{-1}\rho^{2}\partial\rho^{2}-\rho^{2},\\
\end{array}
\end{equation}
and
\begin{equation}
\label{2.7}\begin{array}{c}
a_1=0,\quad b_1=1,\quad c_1=-\frac12u,\\ a_2=\frac14u_x,\quad b_2=-\frac12u,\quad c_2=\frac18(u_{xx}+u^2),\\
a_3=-\frac{1}{16}u_{xxx}-\frac38uu_x,\quad b_3=\frac18u_{xx}+\frac38u^2,\\
c_3=-\frac{1}{32}u_{xxxx}-\frac{7}{32}u_x^2-\frac{3}{16}uu_{xx}-\frac{1}{16}u^3, \\
\end{array}
\end{equation}
where $\partial^{-1}$ is to denote the inverse operator of $\partial$ with the
condition $\partial\partial^{-1}=\partial^{-1}\partial=1$. 

\par
Resorting to the recursive formulae (\ref{2.4}) and (\ref{2.5}), we introduce the Lenard gradients $\{g_j\}\ (j\in\mathbb{Z})$ and the Lenard operator pair $K$, $J$
\begin{equation}\label{2.8}
Kg_{j-1}=Jg_j,\qquad Kg_{-3}=0,\qquad Jg_{-1}=0,\qquad j\in\mathbb{Z},
\end{equation}
where
\begin{equation}\label{2.9}
K=-\frac{1}{2}\partial^3-(\partial u+u\partial),\quad J=2\partial,\quad g_j=-b_{j+2}.
\end{equation}
It is noticed that
\begin{equation}\label{2.10}
\ker K=\{\varrho_{-1}g_{-3}|\forall \varrho_{-1}\in \mathbb{R}\},\qquad
\ker J=\{\varrho_1g_{-1}|\forall\varrho_1\in \mathbb{R}\},
\end{equation}
up to the $\ker K$, $g_{j-1}=K^{-1}Jg_j\ (j\leq-3)$ gives rise to the negative-order Lenard gradients, while up to the $\ker J$, $g_j=J^{-1}Kg_{j-1}\ (j\geq0)$ leads to the positive-order Lenard gradients. For example, it follows from (\ref{2.6}) and (\ref{2.7}) that the first few members of $\{g_j\}$ read
\begin{equation}\label{2.11}
\begin{array}{c}
g_{-1}=-1,\quad g_{-2}=0,\quad g_{-3}=-\rho^2,\quad g_{-4}=
4\rho^2\partial^{-1}\rho^{-2}\partial^{-1}\rho^{-2}\partial^{-1}\rho^{2}\partial\rho^{2},\\
g_0=\frac12u,\quad g_1=-\frac18u_{xx}-\frac38u^2,\quad
g_2=\frac{1}{32}u_{xxxx}+\frac{5}{32}u_x^2+\frac{5}{16}uu_{xx}+\frac{5}{16}u^3,
\end{array}
\end{equation}
which indicate that the formula (\ref{2.8}) is well-defined, and the coalescence of negative- and positive-order Lenard gradients can be termed by the bidirectional Lenard gradients.

\par
Only for convenience in writing, let us introduce a linear operator by $\sigma:\ \mathbb{R}\longrightarrow sl(2,\mathbb{R})$. It is supposed that the time-dependent $\varphi$ satisfies a linear spectral problem given by the positive-order Lenard gradients
\begin{equation}
\varphi_{t_n}=V^{(n)}\varphi, \qquad n\geq 1,\label{2.12}
\end{equation}
where
\begin{equation}V^{(n)}=\sigma(u,\lambda)[g_+]=\frac{1}{2}
\partial g_+\sigma_1-g_+\sigma_2+(\frac{1}{2}\partial^2g_{+}+\lambda
g_++ug_+)\sigma_3,
\label{2.13}
\end{equation}
\begin{equation}
\label{2.14}
g_+=\sum^n_{j=0}g_{j-2}\lambda^{n-j}.
\end{equation}
With the isospectral nature $\lambda_{t_n}=0$, the zero-curvature equation
$$U_{t_n}-V^{(n)}_x+[U,V^{(n)}]=0,$$
yields the usual KdV (pKdV) hierarchy
\begin{equation}
u_{t_n}=Jg_{n-1}\triangleq X_{n-1},\qquad n\geq
1,\label{2.15}\end{equation}
together with a fundamental identity
\begin{equation}V^{(n)}_x-[U,V^{(n)}]=U_*((K-\lambda J)g_+),\label{2.16}
\end{equation}
where $U_*(\xi)=\left.\frac{d}{d\varepsilon}\right|_{\varepsilon=0}U(u+\varepsilon
\xi).$
It is obvious to see that the first nontrivial member in (\ref{2.15}) is the KdV equation
\begin{equation}
u_{t_2}=-\frac{1}{4}(u_{xxx}+6uu_x), \label{2.17}\end{equation} which invokes the Lax representations (\ref{2.1}) and
\begin{equation}\varphi_{t_2}=V^{(2)}\varphi,\qquad
V^{(2)}=\left(\begin{array}{cc}\frac14 u_x&\lambda-\frac12 u\\
-\lambda^2-\frac12 \lambda u+\frac14 u_{xx}+\frac12 u^2&-\frac14 u_x
\end{array}\right).\label{2.18}\end{equation}

\par
On the other hand, it is assumed that the time-dependent $\varphi$ also obeys another linear spectral problem determined by the negative-order Lenard gradients
\begin{equation}
\varphi_{t_{-n}}=V^{(-n)}\varphi, \qquad n\geq 2,\label{2.19}
\end{equation}
where
\begin{equation}V^{(-n)}=\sigma(u,\lambda)[g_-]=\frac{1}{2}
\partial g_-\sigma_1-g_-\sigma_2+(\frac{1}{2}\partial^2g_{-}+\lambda
g_-+ug_-)\sigma_3,
\label{2.20}
\end{equation}
\begin{equation}
\label{2.21}
g_-=-\sum^{n-1}_{j=1}g_{-j-2}\lambda^{j-n}.
\end{equation}
In a similar way, the compatibility of Lax representations of (\ref{2.1}) and (\ref{2.19}) leads to the second fundamental identity
\begin{equation}
V^{(-n)}_x-[U,V^{(-n)}]=U_*[(K-\lambda J)g_{-}], \label{2.22}
\end{equation}
and a hierarchy of nonlocal INLEEs, namely the nKdV hierarchy \cite{4}
\begin{equation}
u_{t_{-n}}=Jg_{-n-1}\triangleq X_{-n+1},\qquad n\geq2, \label{2.23}
\end{equation}
which is characterized by the nKdV equation (\ref{1.1}) (or (\ref{1.2})) in view of $v=\rho^2$. Specifically, the nKdV equation can be regarded as the compatibility condition of Lax representations (\ref{2.1}) and
\begin{equation}\label{2.24}
\varphi_{t_{-2}}=V^{(-2)}\varphi,\quad V^{(-2)}=\left(\begin{array}{cc}
\rho\rho_x\lambda^{-1}&-\rho^2\lambda^{-1}\\
\rho_x^2\lambda^{-1}+\rho^2&-\rho\rho_x\lambda^{-1}\\
\end{array}\right).
\end{equation}

\par
Based on the above presentations, we have specified the Lax pairs for each equation in the bKdV hierarchy by the bidirectional Lenard gradients, which display their integrability in the sense of Lax compatibility \cite{34}, for the integrability in the sense of bi-Hamiltonian structure, one may refer to the reference \cite{8}. To fix $n=1$ in (\ref{2.12}), it is found that the spectral matrix $V^{(1)}$ is the spectral matrix $U$ delivering the fact that the flow variable $t_1$ is in essential the flow variable $x$. Compared with the spectral matrix $V^{(n)}$, the spectral matrix $V^{(-n)}$ allows negative powers of spectral parameter $\lambda$ with the singularity at $\lambda=0$. Therefore, the INLEEs stayed in the nKdV hierarchy are thus recognized as the negative-order integrable systems.

\setcounter{equation}{0}
\section{Reduction to the nKdV hierarchy}
\label{sec:3}
Let $\lambda_1,\lambda_2,\cdots,\lambda_N$ be $N$ distinct nonzero eigenvalues, and $p_j$, $q_j$ be a pair of eigenfunctions associated with $\lambda_j$, $(1\leq j\leq N)$. Designate $\Lambda={\rm diag}(\lambda_1,\cdots,\lambda_N)$, $p=(p_1,p_2,\cdots,p_N)^T$ and $q=(q_1,q_2,\cdots,q_N)^T$. The diamond bracket $\langle\cdot,\cdot\rangle$ stands for the inner product
in the Euclid space $\mathbb{R}^{N}$; and $\omega^2=dp\wedge dq$ represents the symplectic structure in $\mathbb{R}^{2N}$. The Poisson
bracket of two smooth functions $f=f(p,q)$ and $g=g(p,q)$ is defined by \cite{35}
\begin{equation}\label{3.1}\{f,g\}=\sum\limits_{j=1}^N\left(\frac{\pa f}{\pa q_j}\frac{\pa g}{\pa
p_j}-\frac{\pa f}{\pa p_j}\frac{\pa g}{\pa q_j}\right)=\left\al\frac{\pa f}{\pa
q},\frac{\pa g}{\pa p}\right\ar-\left\al\frac{\pa f}{\pa p},\frac{\pa g}{\pa
q}\right\ar,\end{equation} whose value is the derivative of $f$ along with the $g$-flow in
the symplectic space $(\mathbb{R}^{2N},\omega^2)$. We consider $N$ copies of linear spectral problem (\ref{2.1})
\begin{equation}\left(\begin{array}{c}
p_j\\
q_j\\
\end{array}\right)_x=\left(\begin{array}{cc}
0&1\\
-\lambda_j-u&0\\
\end{array}\right)\left(\begin{array}{c}
p_j\\
q_j\\
\end{array}\right) ,\qquad 1\leq j\leq N,\label{3.2}
\end{equation}
which results in the formulae
\begin{equation}\nabla \lambda_j=\delta\lambda_j/\delta u=-p_j^2,
\label{3.3}
\end{equation}
\begin{equation}
(K-\lambda_j J)\nabla \lambda_j=0, \label{3.4}
\end{equation}
\begin{equation}
\sigma(u,\lambda_j)[\nabla \lambda_j]=-p_jq_j\sigma_1+p_j^2\sigma_2-q_j^2\sigma_3\triangleq {\epsilon}_j,
\label{3.5}
\end{equation}
\begin{equation}
\pa{\epsilon}_j=[U(u,\lambda_j),{\epsilon}_j], \label{3.6}
\end{equation}
where $\nabla \lambda_j$ is the functional gradient of $\lambda_j$ with respect to the spectral potential $u$.

\par
From the nonlinearization of Lax pair \cite{25}, we turn to the symmetric (or Neumann) constraint
\begin{equation}
g_{-1}=\sum\limits^N_{j=1}\nabla\lambda_j,\label{3.7}
\end{equation}
and then achieve two geometric conditions
\begin{equation}
 \langle p,p\rangle=1,\qquad \langle p,q\rangle=0, \label{3.8}
\end{equation}
as well as a Neumann map from the eigenfunctions $p$ and $q$ to the spectral
potential $u$
\begin{equation} u=\langle q,q\rangle-\langle\Lambda p,p\rangle.\label{3.9}
\end{equation}
\begin{lemma}\label{lem:1}
\begin{equation}\label{3.10}
(\partial^3+2(u\partial+\partial u))\langle\Lambda^{-k-1}p,p\rangle+4\partial \langle\Lambda^{-k}p,p\rangle=0,\qquad k\geq0.
\end{equation}
\end{lemma}
{\bf Proof:} By using equations (\ref{3.2}), (\ref{3.8}), and (\ref{3.9}), the identity (\ref{3.10}) can be confirmed through a direct calculation, which plays an important role in the reduction of nKdV hierarchy.

\par
To set $k=0$ in (\ref{3.10}), it is clear to see that
\begin{equation}
\label{3.11}(\partial^3+2(u\partial+\partial u))\langle\Lambda^{-1}p,p\rangle=0.\end{equation}
The combination of (\ref{2.8}), (\ref{2.9}), (\ref{2.11}), (\ref{3.10}) and (\ref{3.11}) gives
\begin{equation}
\label{3.12}
\rho=\sqrt{\langle\Lambda^{-1}p,p\rangle}, \qquad g_{-j}=-\langle\Lambda^{-j+2}p,p\rangle,
\qquad j\geq3.
\end{equation}
Moreover, using (\ref{3.2}), (\ref{3.8}), and (\ref{3.9}), a direct calculation results in
\begin{equation}\label{3.12-1}
(\partial^2+u)\rho=-\rho^{-3}(\langle\Lambda^{-1}p,p\rangle-\langle\Lambda^{-1}p,p\rangle\langle\Lambda^{-1}q,q\rangle+
\langle\Lambda^{-1}p,q\rangle^2),
\end{equation}
which confirms that $\rho$ is the kernel of Schr\"{o}dinger operator $\partial^2+u$ as the constant of motion $F_{-0}$ being zero (see (\ref{3.36}) below). Note that the potential $\rho$ can be represented as the square eigenfunction. The dynamical system (\ref{3.12-1}) becomes the Mel'nikov flow originally proposed in \cite{melnikov}, which has been known to be integrable under the inverse scattering transformation.

Note that the pKdV hierarchy (\ref{2.15}) is depicted as the compatibility condition of spectral problems (\ref{2.1}) and (\ref{2.12}), and the nKdV hierarchy (\ref{2.23}) is represented as the compatibility condition of spectral problems (\ref{2.1}) and (\ref{2.19}). Substituting (\ref{3.8}), (\ref{3.9}), and (\ref{3.12}) into (\ref{2.1}), (\ref{2.18}), (\ref{2.24}), and (\ref{2.19}), respectively, we arrive at the classical Neumann system \cite{20,21}
\begin{equation}\label{3.13}
\left\{\begin{array}{l} p_x=q,\\
q_x=-\Lambda p+(\langle\Lambda p,p\rangle-\langle q,q\rangle)p,\\
\langle p,p\rangle=1,\quad \langle p,q\rangle=0,
\end{array}\right.
\end{equation}
and a series of nonlinear dynamical systems of ordinary differential equations
\begin{equation}\label{3.14}
\left\{\begin{array}{l} p_{t_2}=-\langle\Lambda p,q\rangle p+\Lambda q-\frac12\langle q,q\rangle q+\frac12\langle\Lambda p,p\rangle q,\\
q_{t_2}=-\Lambda^2p+\frac12(\langle\Lambda p,p\rangle-\langle
q,q\rangle)\Lambda p+(\langle\Lambda^2p,p\rangle\\-\langle\Lambda
q,q\rangle
-\frac12\langle\Lambda p,p\rangle^2+\frac12 \langle q,q\rangle^2)p+\langle\Lambda p,q\rangle q,\\
\langle p,q\rangle=0,\quad \langle p,p\rangle=1,\\
\end{array}\right.
\end{equation}
\begin{equation}\label{3.15}
\left\{\begin{array}{lll}
p_{t_{-2}}&=&\langle\Lambda^{-1}p,q\rangle\Lambda^{-1}p-\langle\Lambda^{-1}p,p\rangle\Lambda^{-1}q,\\
q_{t_{-2}}&=&\langle\Lambda^{-1}q,q\rangle\Lambda^{-1}p-\Lambda^{-1}p+\langle\Lambda^{-1}p,p\rangle p-
\langle\Lambda^{-1}p,q\rangle\Lambda^{-1}q,\\
&&\langle p,q\rangle=0,\quad \langle p,p\rangle=1,\\
\end{array}\right.
\end{equation}
\begin{equation}\label{3.16}
\left\{\begin{array}{lll}
p_{t_{-k-2}}&=&\frac12\sum\limits_{j=0}^k(\langle\Lambda^{-j-1}p,q\rangle\Lambda^{-k+j-1}p+
\langle\Lambda^{-k+j-1}p,q\rangle\Lambda^{-j-1}p\\&&-2\langle\Lambda^{-j-1}p,p\rangle\Lambda^{-k+j-1}q),\\
q_{t_{-k-2}}&=&-\Lambda^{-k-1}p+\langle\Lambda^{-k-1}p,p\rangle p+\frac12\sum\limits_{j=0}^k
(2\langle\Lambda^{-k+j-1}q,q\rangle\Lambda^{-j-1}p\\&&-\langle\Lambda^{-j-1}p,q\rangle\Lambda^{-k+j-1}q-
\langle\Lambda^{-k+j-1}p,q\rangle\Lambda^{-j-1}q),\\
&&\langle p,q\rangle=0,\quad \langle p,p\rangle=1,\\
\end{array}\right.\quad k\geq1.
\end{equation}
{\bf Note:} Since the time-dependent Lax representation (\ref{2.19}) corresponds to the negative direction of Schr\"{o}dinger spectral problem (\ref{2.1}), the nonlinear dynamical systems (\ref{3.15}) and (\ref{3.16}) are thus called the backward Neumann systems; likewise, due to the Lax representation (\ref{2.12}) staying in the positive direction and the fact of $U=V^{(1)}$, the nonlinear dynamical systems (\ref{3.13}) and (\ref{3.14}) are congruously termed by the forward Neumann systems.

\par
Have a look on the profile of dynamical systems (\ref{3.13})-(\ref{3.16}), it is obvious to see that the backward and forward Neumann systems are all provided with the common geometric condition (\ref{3.8}), which restricts themselves onto the tangent bundle of unit sphere
\begin{equation}\label{3.17}
TS^{N-1}=\{\langle p,q\rangle\in \mathbb{R}^{2N}| F\triangleq \langle
p,q\rangle=0,\ G\triangleq \langle p,p\rangle-1=0\},
\end{equation}
where $F$ and $G$ are two Casimir functions.
Thus, we prefer to use the Dirac-Poisson bracket
\begin{equation}\label{3.18}
\{f,g\}_{D}=\{f,g\}+\frac{1}{\{F,G\}}(\{f,F\}\{G,g\}-\{f,G\}\{F,g\}),
\end{equation}
whose value is the directional derivative of $f$ in the direction of $g$-flow on $(\omega^2,TS^{N-1})$.
With the Dirac-Poisson bracket (\ref{3.18}), the backward and forward Neumann systems (\ref{3.13})-(\ref{3.16}) are represented as the canonical Hamiltonian equations
\begin{equation}\label{3.19}
p_x=\{p, H_1\}_{D},\qquad q_x=\{q,
H_1\}_{D},
\end{equation}
\begin{equation}\label{3.20}
p_{t_2}=\{p, H_2\}_{D},\qquad q_{t_2}=\{q,
H_2\}_{D},
\end{equation}
\begin{equation}\label{3.21}
p_{t_{-2}}=\{p, H_{-0}\}_{D},\qquad q_{t_{-2}}=\{q,
H_{-0}\}_{D},
\end{equation}
\begin{equation}\label{3.22}
p_{t_{-k-2}}=\{p, H_{-k}\}_{D},\quad q_{t_{-k-2}}=\{q,
H_{-k}\}_{D},\quad k\geq1,
\end{equation}
where
\begin{equation}\label{3.23}
H_1=-\frac12(\langle\Lambda p,p\rangle+\langle q,q\rangle),
\end{equation}
\begin{equation}\label{3.24}
H_2=-\frac12(\langle\Lambda^2p,p\rangle+\langle\Lambda
q,q\rangle)-\frac14\langle\Lambda
p,p\rangle\langle q,q\rangle+\frac18(\langle\Lambda p,p\rangle^2+\langle q,q\rangle^2),
\end{equation}
\begin{equation}\label{3.25}
H_{-0}=-\frac12(\langle\Lambda^{-1}p,p\rangle-\langle\Lambda^{-1}p,p\rangle\langle\Lambda^{-1}q,q\rangle+
\langle\Lambda^{-1}p,q\rangle^2),
\end{equation}
\begin{equation}\label{3.26}
H_{-k}=-\frac12\langle\Lambda^{-k-1}p,p\rangle+\frac12\sum\limits_{j=0}^k(
\langle\Lambda^{-j-1}p,p\rangle\langle\Lambda^{-k+j-1}q,q\rangle-\langle\Lambda^{-j-1}p,q\rangle
\langle\Lambda^{-k+j-1}p,q\rangle).
\end{equation}

\par
 Resorting to a special solution of Lenard eigenvalue equation \cite{36}
\begin{equation}\label{3.27}
(K-\lambda J)G_\lambda=0,\quad G_\lambda=-\sum\limits_{j=1}^N\frac{p_j^2}{\lambda-\lambda_j}\triangleq-Q_\lambda(p,p),
\end{equation}
we derive a Lax matrix of the Neumann system (\ref{3.13})
\begin{equation}\label{3.28}
V_\lambda=-Q_\lambda(p,q)\sigma_1+Q_\lambda(p,p)\sigma_2-(1+Q_\lambda(q,q))\sigma_3,
\end{equation}
which satisfies the Lax equation
\begin{equation}\label{3.29}
(V_\lambda)_x-[U,V_\lambda]=0.
\end{equation}
The determinant of $V_\lambda$, under $|\lambda|>\max\{|\lambda_1|,|\lambda_2|,\cdots,|\lambda_N|\}$, results in the usual (forward) integrals of motion to the Neumann system (\ref{3.13}) \cite{37}
\begin{equation}\label{3.30}
F^{+}_\lambda\triangleq \det V_\lambda=\sum\limits_{j=1}^N\frac{E_j}{\lambda-\lambda_j}
=\lambda^{-1}+\sum\limits_{k=1}^\infty F_{k}\lambda^{-k-1},
\end{equation}
where
\begin{eqnarray}
F_1&=&\langle\Lambda p,p\rangle+\langle q,q\rangle,\label{3.31}\\
F_2&=&\langle\Lambda^2p,p\rangle+\langle\Lambda p,p\rangle\langle q,q\rangle+\langle\Lambda q,q\rangle,\label{3.32}\\
F_k&=&\langle\Lambda^kp,p\rangle+\sum\limits_{j=0}^{k-1}\left|\begin{array}{cc}
\langle\Lambda^jp,p\rangle&\langle\Lambda^{k-1-j}p,q\rangle\\
\langle\Lambda^jp,q\rangle&\langle\Lambda^{k-1-j}q,q\rangle \\
\end{array}\right|,\quad
 k\geq 3,\label{3.33}
\end{eqnarray}
and
\begin{equation}\label{3.34}
E_j=p_j^2+\sum\limits_{k=1,k\neq j}^N\frac{(p_jq_k-p_kq_j)^2}{\lambda_j-\lambda_k},
\end{equation}
while in the case of $|\lambda|<\min\{|\lambda_1|,|\lambda_2|,\cdots,|\lambda_N|\}$ leads to the backward integrals of motion
\begin{equation}\label{3.35}
F^{-}_\lambda\triangleq \det V_\lambda=-\sum\limits_{k=0}^\infty\left(\sum\limits_{j=1}^N\lambda_j^{-k-1}E_j\right)\lambda^{k}=-\sum\limits_{k=0}^\infty F_{-k}\lambda^{k},
\end{equation}
\begin{eqnarray}
F_{-0}&=&\langle\Lambda^{-1}p,p\rangle-\langle\Lambda^{-1}p,p\rangle\langle\Lambda^{-1}q,q\rangle+
\langle\Lambda^{-1}p,q\rangle^2,\label{3.36}\\
F_{-1}&=&\langle\Lambda^{-2}p,p\rangle(1-\langle\Lambda^{-1}q,q\rangle)
-\langle\Lambda^{-1}p,p\rangle\langle\Lambda^{-2}q,q\rangle+
2\langle\Lambda^{-1}p,q\rangle\langle\Lambda^{-2}p,q\rangle,\label{3.37}\\
F_{-k}&=&\langle\Lambda^{-k-1}p,p\rangle-\sum\limits_{j=0}^k(
\langle\Lambda^{-j-1}p,p\rangle\langle\Lambda^{-k+j-1}q,q\rangle-\langle\Lambda^{-j-1}p,q\rangle
\langle\Lambda^{-k+j-1}p,q\rangle).\label{3.38}
\end{eqnarray}
With the forward integrals of motion (\ref{3.31})-(\ref{3.33}), we have introduced a sequence of forward Neumann systems \cite{36}
\begin{equation}
p_{t_k}=\{p,H_k\}_{D},\qquad q_{t_k}=\{q,H_k\}_{D},\qquad
k\geq3, \label{3.39}\end{equation}
where
\begin{equation}\label{3.40}
H_k=-\frac12F_k+\frac12\sum_{j=0}^kH_jH_{k-j},\qquad k\geq3,
\end{equation}
which is in agreement with
\begin{equation}\label{3.41}
F^+_\lambda=\lambda^{-1}(1-H^+_\lambda)^2,\quad H^+_\lambda=\sum_{k=0}^\infty H_k\lambda^{-k},
\end{equation}
in view of a supplementary definition $H_0=0$. It has been known that the forward Neumann systems (\ref{3.13}), (\ref{3.14}) and (\ref{3.39}) constitute the decomposition of pKdV hierarchy (\ref{2.15}) \cite{36}. It will be seen that the Neumann systems (\ref{3.13}) and the backward Neumann systems (\ref{3.15}) and (\ref{3.16}) exactly contributes to the decomposition of nKdV hierarchy (\ref{2.23}), see section \ref{sec:5} below.

\setcounter{equation}{0}
\section{The Liouville integrability of backward Neumann systems}
\label{sec:4}

It turns out that the forward Neumann flows commute with each other, completely integrable in the Liouville sense, sharing the same set of integrals of motion: $H_1,H_2,\cdots,H_{N-1},$ functionally independent and involutive in pairwise \cite{36}. This section is to focus on the Liouville integrability of backward Neumann systems (\ref{3.15}) and (\ref{3.16}). In order to handle all the backward Neumann systems synchronously, we would like to use the generating function method. Recalling (\ref{3.25}), (\ref{3.26}), and (\ref{3.36})-(\ref{3.38}), it is clear to see that $H_{-k}=-\frac12 F_{-k},\ k\geq0$. We bring in a generating function for the Hamiltonians $\{H_{-k}\}$
\begin{equation}
H^-_\lambda=-\frac12F^-_\lambda,\qquad
H^-_\lambda=\sum_{k=0}^\infty H_{-k}\lambda^k.\label{4.1}\end{equation}
Let $\tau^+_\lambda$, $\tau^-_\lambda$, $\tau_k \ (k\geq1)$, $\tau_{-k} \ (k\geq0)$, $t^+_\lambda$, $t^-_\lambda$, $t_{k} \ (k\geq1)$ and $t_{-k-2}\ (k\geq0)$ be the flow variables of $F^+_\lambda$, $F^-_\lambda$, $F_k \ (k\geq1)$, $F_{-k} \ (k\geq0)$, $H^+_\lambda$, $H^-_\lambda$, $H_k \ (k\geq1)$ and $H_{-k}\ (k\geq0)$, respectively. It follows from the definition of $F^\pm_\lambda$ in (\ref{3.30}) and (\ref{3.35}) that the canonical Hamiltonian equation (see Eq. (3.23) in \cite{36}) can be extended to
\begin{equation}
\frac{d}{d\tau^{\pm}_\lambda}\left(\begin{array}{c}
p_k\\
q_k\\
\end{array}\right)
=\left(\begin{array}{c}
\{p_k,F^{\pm}_\lambda\}_{D}\\
\{q_k,F^{\pm}_\lambda\}_{D}\\
\end{array}\right)=W(\lambda,\lambda_k)\left(\begin{array}{c}
p_k\\
q_k\\
\end{array}\right) ,\label{4.2}
\end{equation}
where
\begin{equation}
W(\lambda,\mu)=-\frac{2}{\lambda-\mu}V_\lambda-2Q_\lambda(p,p)\sigma_3.
\label{4.3}\end{equation}
Furthermore, on $(TS^{N-1},\omega^2)$ we have the generalized Lax equation
\begin{equation}
\frac{dV_\mu}{d\tau^{\pm}_\lambda}=[W(\lambda,\mu)
, V_\mu],\qquad \lambda\neq\mu,\quad \lambda,\ \mu\in\mathbb{C}, \label{4.4}
\end{equation}
which gives rise to the formulae
\begin{equation}\label{4.5}
\{F^{\pm}_\mu,F^{\pm}_\lambda\}_{D}=0,\qquad \lambda\neq\mu,\quad \forall \lambda,\ \mu\in\mathbb{C},
\end{equation}
\begin{equation}\label{4.6}
\{H^{\pm}_\mu,H^{\pm}_\lambda\}_{D}=0,\qquad \lambda\neq\mu,\quad\forall \lambda,\
\mu\in\mathbb{C},
\end{equation}
and by inserting (\ref{3.30}), (\ref{3.35}), $(\ref{3.41})_2$ and $(\ref{4.1})_2$ into (\ref{4.5}) and (\ref{4.6}) leads to the involutivity for all kinds of integrals of motion $\{F_k,\ k\geq1\}\cup\{F_{-k},\ k\geq0\}\cup\{H_k,\ k\geq1\}\cup\{H_{-k},\ k\geq0\}$
\begin{equation}\label{4.7}
\{F_{\pm j},F_{\pm k}\}_{D}=0,\qquad j,\ k=0,1,2,\cdots,
\end{equation}
\begin{equation}\label{4.8}
\{H_{\pm j},H_{\pm k}\}_{D}=0,\qquad j,\ k=0,1,2,\cdots.
\end{equation}
In addition, by the definition of Dirac-Poisson bracket, a direct calculation gives
\begin{equation}\label{4.9}
\frac{dF_\mu^{\pm}}{dt_\lambda^{-}}=\{F_\mu^{\pm},H_\lambda^{-}\}_D=-\frac12\{F_\mu^{\pm},F_\lambda^{-}\}_D=0,
\end{equation}
\begin{equation}\label{4.10}
\frac{dF^{\pm}_\mu}{dt^{+}_\lambda}=\left\{\begin{array}{l}\{F^{+}_\mu,H^{+}_\lambda\}_{D}=
-2\mu^{-1}(1-H^{+}_\mu)\{H^{+}_\mu,H^{+}_\lambda\}_{D}=0,\\
\{F^{-}_\mu,H^{+}_\lambda\}_{D}=-2\{H_\mu^{-},H^{+}_\lambda\}_D=0,\\
\end{array}\right.
\end{equation}
which together with (\ref{4.6}) imply that, $\{F_{k}\}$ and $\{H_{k}\}$ $(k\in\mathbb{Z})$, are integrals of motion for all the backward and forward Neumann systems (\ref{3.13})-(\ref{3.16}) and (\ref{3.39}).

\par
According to the Liouville's definition, the other essential element to the integrability is the functional independence of integrals of motion. The Hamiltonians $H_1,H_2,\cdots,H_{N-1},$ have been shown to be functionally independent via the $\epsilon$-technique of an algebraic calculation \cite{38}. The rest of this section is to specify the functional independence of backward Hamiltonians $H_{-0},H_{-1},\cdots,H_{-N+2}$ by using a set of quasi-Abel-Jacobi variables in the context of algebraic geometry. For the brevity of presentations, let us make the symbols
\begin{equation}
\label{4.11}
F^\pm_\lambda=-V_\lambda^{12}V_\lambda^{21}-(V_\lambda^{11})^2,\qquad V_\lambda=
V_\lambda^{11}\sigma_1+V_\lambda^{12}\sigma_2+V_\lambda^{21}\sigma_3.
\end{equation}
To progress further, we define
\begin{equation}
\label{4.12}
F^\pm_\lambda=\frac{b(\lambda)}{a(\lambda)}=\frac{R(\lambda)}{a^2(\lambda)},
\end{equation}
\begin{equation}
\label{4.13} V_\lambda^{12}=Q_\lambda(p,p)=\frac{m(\lambda)}{a(\lambda)},
\end{equation}
where
\begin{equation}
\label{4.14}\begin{array}{lll}
a(\lambda)&=&\prod\limits_{k=1}^N(\lambda-\lambda_k),\quad b(\lambda)=\prod\limits_{k=1}^{N-1}(\lambda -\lambda_{N+k}),\\
R(\lambda)&=&\prod\limits_{k=1}^{2N-1}(\lambda-\lambda_k),\quad
m(\lambda)=\prod\limits_{k=1}^{N-1}(\lambda-\mu_k),\\
\end{array}\end{equation}
and $\mu_1,\mu_2,\cdots,\mu_{N-1}$ are $N-1$
elliptic variables of the backward and forward Neumann systems (\ref{3.13})-(\ref{3.16}) and (\ref{3.39}).

\par
Substituting $\lambda$ with $\mu_k$ in $(\ref{4.11})_1$ and (\ref{4.12}), we have
\begin{equation}
\label{4.15}
\left.V_\lambda^{11}\right|_{\lambda=\mu_j}=\frac{\sqrt{-R(\mu_j)}}{a(\mu_j)},\qquad 1\leq j\leq N-1.
\end{equation}
By virtue of (\ref{4.3}), (\ref{4.4}), (\ref{4.13}) and (\ref{4.15}), we arrive at the Dubrovin type equation
\begin{equation}\label{4.16}
\frac{1}{4\sqrt{-R(\mu_k)}}\frac{d\mu_k}{d\tau^{\pm}_\lambda}=-\frac{m(\lambda)}{a(\lambda)(\lambda-\mu_k)m'(\mu_k)},\qquad 1\leq k\leq N-1,
\end{equation}
that governs the evolution of $\mu_k$ along with the $\tau^{\pm}_\lambda$-flow on $(TS^{N-1},\omega^2)$. Multiplying (\ref{4.16}) by $\mu_k^{N-1-j}$ and summing with regard to $k=1,2,\cdots,N-1$, we derive
\begin{equation}\label{4.17}
\sum\limits_{k=1}^{N-1}\frac{\mu_k^{N-1-j}}{4\sqrt{-R(\mu_k)}}\frac{d\mu_k}{d\tau^{\pm}_\lambda}
=-\frac{\lambda^{N-1-j}}{a(\lambda)},\qquad 1\leq j\leq N-1,
\end{equation}
with the aid of the Lagrange interpolation formula. The stimulus for considering the hyperelliptic curve of Riemann surface $\Gamma$
\begin{equation}\label{4.18}
\xi^2+R(\lambda)=0,
\end{equation}
is hence inspired by the shape of (\ref{4.17}). Due to $\deg R(\lambda)=2N-1$, the Riemann surface $\Gamma$ is of genus $N-1$, and there exists only one infinity that is the branch point of $\Gamma$. As for a given $\lambda \ (\neq\infty,\lambda_k,k=1,2,\cdots,2N-1)$, there are two points $P_{\pm}(\lambda)=(\lambda,\pm\sqrt{-R(\lambda)})$ on the upper and lower sheets of $\Gamma$; specially, to set $\lambda=0$, the corresponding two points are expressed by $0_1=(0,-\sqrt{-R(0)})$ and $0_2=(0,\sqrt{-R(0)})$. On the Riemann surface $\Gamma$, the $N-1$ linearly independent holomorphic differentials are
\begin{equation}\label{4.19}
\tilde{\omega}_j=\frac{\lambda^{N-1-j}d\lambda}{4\sqrt{-R(\lambda)}},\quad\quad
\quad 1\leq j\leq N-1.
\end{equation}
Denote a fixed point $P_0 \ (\neq\infty,0_1,0_2,\lambda_j\ (j=1,2,\cdots,2N-1))$ on $\Gamma$. Let us now introduce a set of quasi-Abel-Jacobi variables
\begin{equation}\label{4.20}
\tilde{\phi}_j=\sum\limits_{k=1}^{N-1}\int_{P_0}^{P(\mu_k)}\tilde{\omega}_j,\quad\quad
1\leq j\leq N-1.
\end{equation}
With the quasi-Abel-Jacobi variables $\tilde{\phi}_j$, the formula (\ref{4.17}) can be rewritten as
\begin{equation}\label{4.21}
\frac{d\tilde{\phi}_j}{d\tau^{\pm}_\lambda}=
-\frac{\lambda^{N-1-j}}{a(\lambda)},\qquad 1\leq j\leq N-1.
\end{equation}

\par
It follows from the Dirac-Poisson bracket that
\begin{equation}\label{4.22}
\frac{d\tilde{\phi}_j}{d\tau^{-}_\lambda}=\{\tilde{\phi}_j,F_\lambda^{-}\}_D=\sum_{k=0}^\infty\{\tilde{\phi}_j,-F_{-k}\}_D\lambda^k=
-\sum_{k=0}^\infty\frac{d\tilde{\phi}_j}{d\tau_{-k}}\lambda^k, \quad 1\leq j\leq N-1.
\end{equation}
Let $s_{-k}=\sum\limits_{j=1}^{N}\lambda_j^{-k}$ and $s_{k}=\sum\limits_{j=1}^{N}\lambda_j^{k}$. Based on the calculation of series expansions, one has
\begin{equation}\label{4.23}
\frac{1}{a(\lambda)}=\left\{\begin{array}{l}\sum\limits_{k=0}^\infty \mathscr{A}_{-k}\lambda^{k},\quad
|\lambda|<\min\{|\lambda_1|,|\lambda_2|,\cdots,|\lambda_N|\},\\
\sum\limits_{k=0}^\infty \bar{\mathscr{A}}_k\lambda^{-k-N},\quad |\lambda|>\max\{|\lambda_1|,|\lambda_2|,\cdots,|\lambda_N|\},\\
\end{array}\right.
\end{equation}
where
\begin{equation}\label{4.24}\begin{array}{cll}
\mathscr{A}_{k}&=&0 \ (k\geq1),\quad \mathscr{A}_0=(-1)^N\prod_{j=1}^N\lambda_j^{-1},\quad \mathscr{A}_{-1}=\mathscr{A}_0s_{-1},\\
\mathscr{A}_{-2}&=&\frac12\mathscr{A}_0(s_{-2}+s_{-1}^2),\ \mathscr{A}_{-3}=\mathscr{A}_0(\frac13s_{-3}+\frac12s_{-1}s_{-2}+\frac16s_{-1}^3),\\
\mathscr{A}_{-k}&=&\frac1k\left(\mathscr{A}_0s_{-k}+\sum\limits_{i+j=k,i,j\geq1}\mathscr{A}_{-i}s_{-j}\right),\quad
k\geq 4,\\
\end{array}
\end{equation}
and
\begin{equation}\label{x4.24}\begin{array}{lll}
\bar{\mathscr{A}}_{-k}&=&0 \ (k\geq1),\quad \bar{\mathscr{A}}_0=1,\quad \bar{\mathscr{A}}_1=s_1,\\
\bar{\mathscr{A}}_2&=&\frac12(s_2+s_1^2),\quad \bar{\mathscr{A}}_3=\frac16(2s_3+3s_2s_1+s_1^3),\\
\bar{\mathscr{A}}_k&=&\frac1k\left(s_k+\sum\limits_{i+j=k,i,j\geq1}s_i\bar{\mathscr{A}}_j\right),\quad
k\geq 4.\\
\end{array}\end{equation}
Combining (\ref{4.21}) with (\ref{4.23}), we also have
\begin{equation}\label{4.25}
\frac{d\tilde{\phi}_j}{d\tau^{-}_\lambda}=-\sum\limits_{k=0}^\infty \mathscr{A}_{-k}\lambda^{N-1-j+k},\quad 1\leq j\leq N-1.
\end{equation}
The comparison of same powers in (\ref{4.22}) and (\ref{4.25}) yields
\begin{equation}\label{4.26}
\frac{d\tilde{\phi}_j}{d\tau_{-k}}=\mathscr{A}_{N-1-k-j},\qquad 0\leq k\leq N-2,\quad 1\leq j\leq N-1.
\end{equation}
Let $\tilde{\phi}$ be the column vector of $(\tilde{\phi}_1,\tilde{\phi}_2,\cdots,\tilde{\phi}_{N-1})^T$. From (\ref{4.26}), we arrive at
\begin{equation}\label{4.27}
\mathscr{A}\triangleq
\frac{\partial(\tilde{\phi}_1,\tilde{\phi}_2,\cdots,\tilde{\phi}_{N-1})}{\partial(\tau_{-0},\tau_{-1},\cdots,\tau_{-N+2})} =\left(\begin{array}{cccccc}
0&0&0&\cdots&0&\mathscr{A}_0\\
0&0&0&\cdots&\mathscr{A}_0&\mathscr{A}_{-1}\\
\vdots&\vdots&\vdots&\vdots&\vdots&\vdots\\
0&\mathscr{A}_0&\mathscr{A}_{-1}&\cdots&\mathscr{A}_{-N+4}&\mathscr{A}_{-N+3}\\
\mathscr{A}_0&\mathscr{A}_{-1}&\mathscr{A}_{-2}&\cdots&\mathscr{A}_{-N+3}&\mathscr{A}_{-N+2}\\
\end{array}\right).
\end{equation}
\begin{proposition}
 \label{prop:1}
 $\{F_{-0},F_{-1},\cdots,F_{-N+2}\}$
are functionally independent on $(TS^{N-1},\omega^2)$.
\end{proposition}
{\bf Proof:} Followed by \cite{35}, it is sufficient to demonstrate the linear independence of one-form differentials $dF_{-0},dF_{-1},\cdots,dF_{-N+2}$ in the cotangent space $T^{*}_{(p,q)}S^{N-1}$ at any point $(p,q)\in TS^{N-1}$.
Take into account the identity
\begin{equation}\label{4.28}
\gamma_{-0}dF_{-0}+\gamma_{-1}dF_{-1}+\cdots+\gamma_{-N+2}dF_{-N+2}=0,
\end{equation}
where $\gamma_{-0},\gamma_{-1},\cdots,\gamma_{-N+2}$ are $N-1$ constants. Recalling the representation of Poisson bracket in terms of the symplectic structure $\omega^2$, we know that
\begin{equation}\label{4.29}
\sum_{k=0}^{N-2}\gamma_{-k}\{\tilde{\phi}_j,F_{-k}\}_D=\sum_{k=0}^{N-2}\gamma_{-k}\frac{d\tilde{\phi}_j}{d\tau_{-k}}=0.
\end{equation}
Because the coefficient matrix $\mathscr{A}$ is non-degenerate (lower-triangular determinant), one derives
$$\gamma_{-0}=\gamma_{-1}=\cdots=\gamma_{-N+2}=0,$$
which completes the proof.
\begin{corollary}
  \label{coro:1}
$\{H_{-0},H_{-1},\cdots,H_{-N+2}\}$
are functionally independent on $(TS^{N-1},\omega^2)$.
\end{corollary}

\par
To this end, by the involutivity and functional independence of $\{H_{-k}\}$, $k=0,1,2,\cdots,N-2$, we come to the Liouville integrability of backward Neumann systems.
\begin{theorem}
  \label{theo:1}
The backward Neumann systems (\ref{3.15}) and (\ref{3.16}) are completely integrable in the Liouville sense.
\end{theorem}
Based on Proposition 3 in \cite{36} and Theorem \ref{theo:1}, the backward and forward Neumann systems $(H_{-i},\omega^2,\mathbb{R}^{2N})$ and $(H_j,\omega^2,\mathbb{R}^{2N})$ are consistent in pairwise, which indicates that there exists a smooth function associated with flow variables $t_{-i-2}$ and $t_j$ generating the involutivity solution for the backward and forward Neumann systems.

\setcounter{equation}{0}
\section{From the backward Neumann systems to the nKdV hierarchy}
\label{sec:5}
To solve INLEEs by FDISs, one crucial step is to establish the relationship between INLEEs and FDISs \cite{26,27,28,29}. The relationship between the pKdV hierarchy and the forward Neumann systems has been exhibited by means of the Neumann map \cite{36}. This section is devoted to establishing the relationship between the nKdV hierarchy and the backward Neumann systems.

\par
It is known from (\ref{4.9}) and (\ref{4.10}) that $F_1$ is a constant of motion independent of the flow variables. Recalling (\ref{3.9}) and (\ref{3.31}), we have
\begin{equation}\label{5.1}
u=F_1-2\langle\Lambda p,p\rangle,
\end{equation}
from which the spectral potential $u$ can be expressed by the symmetric functions of elliptic variables $\{\mu_k\}$, $k=1,2,\cdots,N-1$.
\begin{lemma}\label{lemm:2}
Let $\lambda\neq\lambda_j,\ (1\leq j\leq N)$. Then
\begin{equation}
\begin{array}{l}
\langle\Lambda p,p\rangle=\sum\limits_{j=1}^N\lambda_j-\sum\limits_{j=1}^{N-1}\mu_j,\\
\langle\Lambda^2 p,p\rangle=\sum\limits_{i<j}\mu_i\mu_j-\sum\limits_{i<j}\lambda_i\lambda_j+(\sum\limits_{j=1}^N\lambda_j)^2-
\sum\limits_{j=1}^N\lambda_j\sum\limits_{j=1}^{N-1}\mu_j,\\
\langle\Lambda^3 p,p\rangle=\sum\limits_{i<j<k}\lambda_i\lambda_j\lambda_k-\sum\limits_{i<j<k}\mu_i\mu_j\mu_k
-2\sum\limits_{j=1}^N\lambda_j\sum\limits_{i<j}\lambda_i\lambda_j\\
+\sum\limits_{j=1}^N\lambda_j\sum\limits_{i<j}\mu_i\mu_j+
(\sum\limits_{i<j}\lambda_i\lambda_j-(\sum\limits_{j=1}^N\lambda_j)^2)\sum\limits_{j=1}^{N-1}\mu_j
+(\sum\limits_{j=1}^N\lambda_j)^3.\\
\end{array}
\label{5.2}\end{equation}
\end{lemma}
{\bf Proof:} With the help of (\ref{4.14}), the equation (\ref{4.13}) is put into the form
\begin{equation}\label{5.3}
\begin{array}{l}\displaystyle{
\sum\limits_{l=1}^Np_l^2(\lambda^{N-1}-\lambda^{N-2}\sum\limits^N_{j=1;j\neq l}\lambda_j
+\cdots+(-1)^{N-1}\prod\limits^N_{j=1;j\neq l}\lambda_j)}\\
\displaystyle{=\lambda^{N-1}-\lambda^{N-2}\sum\limits_{j=1}^{N-1}\mu_j+\cdots+
(-1)^{N-1}\prod\limits^{N-1}_{j=1}\mu_j}.\\
\end{array}
\end{equation}
Comparing the coefficients of $\lambda^{N-2}$, $\lambda^{N-3}$ and $\lambda^{N-4}$ on both sides of (\ref{5.3}), we come to the equation (\ref{5.2}) by virtue of the identities $\sum_{i<j;i,j\neq l}\lambda_i\lambda_j=\sum_{i<j}\lambda_i\lambda_j-\lambda_l\sum_{j=1}^N\lambda_j+\lambda_l^2$
 and $\sum_{i<j<k \atop i,j,k\neq l}\lambda_i\lambda_j\lambda_k=\sum_{i<j<k}\lambda_i\lambda_j\lambda_k-\lambda_l\sum_{i<j}\lambda_i\lambda_j
+\lambda^2_l\sum_{j=1}^N\lambda_j-\lambda_l^3.$

\par
Let us bring in a generating function of the negative-order Lenard gradients
\begin{equation}\label{5.4}
g_\lambda^{-}=\sum\limits_{k=0}^\infty g_{-k-3}\lambda^k,
\end{equation}
which also satisfies the Lenard eigenvalue equation
\begin{equation}\label{5.5}
(K-\lambda J)g_\lambda^{-}=0.
\end{equation}
By applying the operator $K^{-1}J$ on the Neumann constraint (\ref{3.7}) $k$ times, we arrive at
\begin{equation}\label{5.6}
\sum\limits_{j=1}^N\lambda_j^{-k}\nabla\lambda_j=g_{-k-2}+c_{-2}g_{-k-1}+\cdots+c_{-k+1}g_{-4}+c_{-k}g_{-3},\quad k\geq1,
\end{equation}
in view of the kernel of Lenard operator $K$, where $c_{-2},c_{-3},\cdots,c_{-k}$ are constants of integration.
 By $|\lambda|<\min\{|\lambda_1|,|\lambda_2|,\cdots,|\lambda_N|\}$, together with (\ref{5.4}) and (\ref{5.6}) the special solution $G_\lambda$ of Lenard eigenvalue equation can be rewritten as
\begin{equation}\label{5.7}
\begin{array}{lll}
G_\lambda&=&\sum\limits_{j=1}^N\frac{\nabla\lambda_j}{\lambda-\lambda_j}
=-\sum\limits_{k=0}^\infty\left(\sum\limits_{j=1}^N\lambda_j^{-k-1}\nabla\lambda_j\right)\lambda^k\\
&=&-\sum\limits_{k=0}^\infty(g_{-k-3}+c_{-2}g_{-k-2}+\cdots+c_{-k}g_{-4}+c_{-k-1}g_{-3})\lambda^k\\
&=&c_\lambda^{-}g_\lambda^{-},\\
\end{array}
\end{equation}
where
\begin{equation}\label{5.8}
c_\lambda^{-}=-1-\sum\limits_{k=0}^\infty c_{-k-2}\lambda^{k+1}.
\end{equation}
It follows from (\ref{4.1}) that $H_{-k}$ is linear to integrals of motion $F_{-k}$, $k\geq0$. From the definition of Dirac-Poisson bracket, we obtain
\begin{equation}
\label{5.9}
\frac{d}{dt_\lambda^-}=-\frac12\frac{d}{d\tau_\lambda^-},\quad \frac{d}{dt_{-k-2}}=-\frac12\frac{d}{d\tau_{-k}},\qquad k\geq0.
\end{equation}
\begin{proposition}
  \label{prop:2}
Let $(p(x,t_{-k-2}),q(x,t_{-k-2}))^T$ be an involutive solution of the Neumann system $(H_1,TS^{N-1},\omega^2)$ and the backward Neumann system $(H_{-k},TS^{N-1},\omega^2)$, $k\geq0$. Then
\begin{equation}
u=\langle q(x,t_{-k-2}),q(x,t_{-k-2})\rangle-\langle\Lambda p(x,t_{-k-2}),p(x,t_{-k-2})\rangle,
\label{5.10}
\end{equation}
is the finite parametric solution of the $(k+1)$-th nKdV equation (\ref{2.23}).
\end{proposition}
{\bf Proof:} On one hand, by (\ref{3.15}) and (\ref{3.16}), a lengthy but direct calculation gives
\begin{equation}\label{5.11}
\frac{\partial \langle\Lambda p,p\rangle}{\partial t_{-k-2}}=2\langle\Lambda^{-k-1}p,q\rangle,\qquad
\frac{\partial \langle q,q\rangle}{\partial t_{-k-2}}=-2\langle\Lambda^{-k-1}p,q\rangle.
\end{equation}
On the other hand, resorting to (\ref{2.9}), (\ref{3.12}), and (\ref{3.13}), we have
\begin{equation}\label{5.12}
Jg_{-k-3}=-2\partial\langle\Lambda^{-k-1}p,p\rangle=-4\langle\Lambda^{-k-1}p,q\rangle.
\end{equation}
Substituting (\ref{5.11}) and (\ref{5.12}) back into (\ref{2.23}), it turns out that (\ref{5.10}) is a finite parametric solution of the $(k+1)$-th nKdV equation (\ref{2.23}).
\begin{corollary}
  \label{coro:2}
Let $(p(x,t_{-2}),q(x,t_{-2}))$ be an involutive solution of the classical Neumann system $(H_1,TS^{N-1},\omega^2)$ and the backward Neumann system $(H_{-0},TS^{N-1},\omega^2)$. Thus,
\begin{equation}
u=\langle q(x,t_{-2}),q(x,t_{-2})\rangle-\langle\Lambda p(x,t_{-2}),p(x,t_{-2})\rangle,\quad
v=\rho^2=\langle\Lambda^{-1}p(x,t_{-2}),p(x,t_{-2})\rangle,
\label{5.13}
\end{equation}
satisfy the nKdV equation (\ref{1.1}) (or (\ref{1.2})).
\end{corollary}

\par
Based on the commutability of backward and forward Neumann flows, the bKdV hierarchy is reduced to an infinite sequence of backward and forward Neumann systems, especially for the case of nonlocal INLEEs to local FDISs (see Proposition \ref{prop:2}). Therefore, the derivation of explicit solutions for bKdV hierarchy is changed to the problem of solving a family of Neumann systems, which in a sense simplifies the procedure of getting explicit solutions. Within our best knowledge, the key point of the derivation of explicit solutions for INLEEs, no matter pure soliton solutions or quasi-periodic solutions, is to specify a finite-dimensional invariant subspace of IDISs from the infinite-dimensional function space \cite{39,40}. Actually, Novikov had already proposed the conception of high-order stationary KdV (Novikov) equation, which determines a finite-dimensional invariant subspace of KdV flows, namely the solution space of Novikov equation. To solve the nKdV equations, we generalize the Novikov equation to the negative-order case that specifies a finite-dimensional invariant subspace for the nKdV flows, i.e. the negative $N$-order stationary KdV equation.
\begin{theorem}
  \label{theo:2}
It is assumed that $(p(x),q(x))^T$ is a solution of the Neumann system (\ref{3.13}). Thus
\begin{equation}\label{5.14}
 u=\langle q,q\rangle-\langle\Lambda p,p\rangle,
\end{equation}
is the finite-gap potential to the negative-order Novikov (or negative $N$-order stationary KdV) equation
\begin{equation}\label{5.15}
X_{-N}+\bar{c}_{-2}X_{-N+1}+\cdots+\bar{c}_{-N}X_{-1}=0,
\end{equation}
where $\bar{c}_{-j}$ are some constants of integration given by
\begin{equation}\label{5.16}
\bar{c}_{-j}=\sum\limits_{k=0}^{j-1}a_{-k}c_{-j+k},\qquad j=2,3,\cdots,N,
\end{equation}
with a supplementary definition $c_{-1}=1$, and
\begin{equation}\label{5.17}
a_{-0}=1,\qquad a_{-j}=(-1)^j\sum\limits_{i_1<i_2<\cdots<i_j}\lambda_{i_1}^{-1}\lambda_{i_2}^{-1}\cdots\lambda_{i_j}^{-1},\qquad j=1,2,\cdots,N.
\end{equation}
\end{theorem}
{\bf Proof:} With an auxiliary polynomial
\begin{equation}\label{5.18}
a(\lambda^{-1})=\prod_{j=1}^N(\lambda^{-1}-\lambda_j^{-1})=\lambda^{-N}+a_{-1}\lambda^{-N+1}+\cdots+a_{-N+1}\lambda^{-1}+a_{-N},
\end{equation}
we obtain from (\ref{5.6}) that
\begin{equation}\label{5.19}
\begin{array}{ccl}
0&=&\sum\limits_{j=1}^Na(\lambda_j^{-1})\nabla\lambda_j=\sum\limits_{j=1}^N(\lambda_j^{-N}+a_{-1}\lambda_j^{-N+1}
+\cdots+a_{-N+1}\lambda_j^{-1}+a_{-N})\nabla\lambda_j\\
&=&(g_{-N-2}+c_{-2}g_{-N-1}+\cdots+c_{-N+1}g_{-4}+c_{-N}g_{-3})\\
&&+a_{-1}(g_{-N-1}+c_{-2}g_{-N}+\cdots+c_{-N+2}g_{-4}+c_{-N+1}g_{-3})\\
&&+\cdots+a_{-N+2}(g_{-4}+c_{-2}g_{-3})+a_{-N+1}g_{-3}+a_{-N}g_{-1},\\
&=&g_{-N-2}+\bar{c}_{-2}g_{-N-1}+\cdots+\bar{c}_{-N+1}g_{-4}+\bar{c}_{-N}g_{-3}+a_{-N}g_{-1}.\\
\end{array}
\end{equation}
Applying the Lenard operator $J$ on (\ref{5.19}), we attain the negative-order Novikov equation (\ref{5.15}).

\setcounter{equation}{0}
\section{The algebro-geometric construction of nKdV flows}
\label{sec:6}
Subjected to the Neumann type integrable reduction, the KdV and the fifth-order KdV flows are integrated with Abel-Jacobi solutions on the Jacobi variety of a Riemann surface, in which their quasi-periodic solutions are retrieved in view of three forward Neumann systems \cite{36}. In what follows, we make an endeavor to illustrate the evolution behavior of the nKdV flows, and further to write down Riemann theta function representations of the finite-gap potential in terms of backward Neumann systems.

\par
We firstly recollect some algebraic geometrical datum associated with the nKdV flows. Let us introduce a set of canonical basis of homological cycles $\{a_j,b_j\}_{j=1}^{N-1}$ on the Riemann surface $\Gamma$. Taking into account the non-degenerate matrix
\begin{equation}
C=(A_{ij})^{-1}_{N-1\times N-1},\quad A_{ij}=\int_{a_j}\tilde{\omega}_i,
\qquad 1\leq i,j\leq N-1,
\label{6.1}\end{equation}
we arrive at the normalized holomorphic differential
\begin{equation}
\omega=(\omega_1,\omega_2,\cdots,\omega_{N-1})^T,\quad
\omega_j=\sum\limits_{l=1}^{N-1}C_{jl}\tilde{\omega}_l, \qquad 1\leq
j\leq N-1,\label{6.2}\end{equation}
with the property
\begin{equation}
\int_{a_i}\omega_j=
\sum\limits_{l=1}^{N-1}C_{jl}\int_{a_i}\tilde{\omega}_l=\sum\limits_{l=1}^{N-1}C_{jl}A_{li}=
\delta_{ji}=\left\{\begin{array}{ll}
1,& i=j,\\
0,& i\neq j.\\
\end{array}\right.
\label{6.3}\end{equation}
The $2(N-1)$ periodic vectors
\begin{equation}
\delta_j=\int_{a_j}\omega,\qquad B_j=\int_{b_j}\omega,\qquad 1\leq
j\leq N-1, \label{6.4}\end{equation}
span a lattice $\mathcal {T}$ in the complex space $\mathbb{C}^{N-1}$, which defines the Jacobi variety
$J(\Gamma)=\mathbb{C}^{N-1}/\mathcal {T}$ of $\Gamma$.
Let $\delta=(\delta_1,\delta_2,\cdots,\delta_{N-1})$ and $B=(B_1,B_2,\cdots,B_{N-1})$. It is seen from (\ref{6.3}) that $\delta$ is a unit matrix. By the Riemannian bilinear relation, $B$ is a symmetric matrix with positive-definite imaginary part, and can also be used to construct the Riemann theta function of $\Gamma$ \cite{41,42}
\begin{equation}\theta(\varsigma)=\sum\limits_{z\in \mathbb{Z}^{N-1}}\exp{\pi\sqrt{-1}(
\langle Bz,z\rangle+2\langle\varsigma,z\rangle)},\qquad \varsigma\in \mathbb{C}^{N-1}.
\label{6.5}\end{equation}

\par
The Abel map $\mathcal {A}:\ {\rm Div}(\Gamma)\longrightarrow J(\Gamma)$ is defined from the divisor group to the Jacobi variety
\begin{equation}\label{6.6}
\mathcal {A}(P)=\int_{P_0}^P\omega,\qquad \mathcal
{A}\left(\sum_{k=1}^{N-1} n_kP_k\right)=\sum_{k=1}^{N-1} n_k\mathcal
{A}(P_k).
\end{equation}
Taking $\sum_{k=1}^{N-1}P(\mu_k)$ as a special divisor, on $J(\Gamma)$ we elaborate the Abel-Jacobi variable
\begin{equation}\label{6.7}
\bar{\phi}=C\tilde{\phi}={\cal{A}}\left(\sum\limits_{k=1}^{N-1}P(\mu_k)\right)=\sum\limits_{k=1}^{N-1}\int_{P_0}^{P(\mu_k)}\omega,
\end{equation}
where $$P(\mu_k)=(\mu_k,\xi(\mu_k)),\qquad C=(C_1,C_2,\cdots,C_{N-1}).$$

\begin{lemma}\label{lemm:3}
Let $S_{-k}=\sum\limits_{j=1}^{2N-1}\lambda_j^{-k}$ and $S_{k}=\sum\limits_{j=1}^{2N-1}\lambda_j^{k}$. Then
\begin{equation}\label{6.8}
\frac{1}{\sqrt{R(\lambda)}}=\left\{\begin{array}{l}\sum\limits_{k=0}^\infty\tilde{\Lambda}_{-k}\lambda^k,\qquad |\lambda|<\min\{|\lambda_1|,|\lambda_2|,\cdots,|\lambda_{2N-1}|\},\\
\sum\limits_{k=0}^\infty\bar{\Lambda}_{k}\lambda^{-k-N+\frac12},\qquad |\lambda|>\max\{|\lambda_1|,|\lambda_2|,\cdots,|\lambda_{2N-1}|\},\\
\end{array}\right.
\end{equation}
where
\begin{equation}\label{6.9}\begin{array}{cll}
\tilde{\Lambda}_k&=&0,\ \ (k\geq1),\quad \tilde{\Lambda}_0=\left(-\prod\limits_{j=1}^{2N-1}\lambda_j\right)^{-\frac12},\quad \tilde{\Lambda}_{-1}=\frac12\tilde{\Lambda}_0S_{-1},\\
\tilde{\Lambda}_{-2}&=&\frac18\tilde{\Lambda}_0(2S_{-2}+S_{-1}^2),\ \tilde{\Lambda}_{-3}=\frac{1}{48}\tilde{\Lambda}_0(8S_{-3}+6S_{-1}S_{-2}+S_{-1}^3),\vspace{2mm}\\
\tilde{\Lambda}_{-k}&=&\frac{1}{2k}(\tilde{\Lambda}_0S_{-k}+\sum\limits_{i+j=k;i,j\geq1}S_{-i}\tilde{\Lambda}_{-j}),\qquad k\geq4,\\
\end{array}
\end{equation}
and
\begin{equation}\label{6.10}\begin{array}{lll}
\bar{\Lambda}_{-k}&=&0,\ \ (k\geq1),\quad \bar{\Lambda}_0=1,\quad \bar{\Lambda}_{1}=\frac12S_{1},\\
\bar{\Lambda}_{2}&=&\frac18(2S_{2}+S_{1}^2),\ \bar{\Lambda}_{3}=\frac{1}{48}(8S_{3}+6S_{1}S_{2}+S_{1}^3),\\
\bar{\Lambda}_{k}&=&\frac{1}{2k}(S_{-k}+\sum\limits_{i+j=k;i,j\geq1}S_{i}\bar{\Lambda}_{j}),\qquad k\geq4.\\
\end{array}
\end{equation}
\end{lemma}
{\bf Proof:} The proof is based on a direct calculation of power series expansions.

\par
After these preparations, it is remarkable to see that the Abel-Jacobi variable $\bar{\phi}$ straightens out not only the backward Neumann flows, but also the forward Neumann flows.
\begin{theorem}
  \label{theo:3}
The backward Neumann flows are linearized by the Abel-Jacobi variable $\bar{\phi}$ on $J(\Gamma)$
\begin{equation}\label{6.11}
\frac{d\bar{\phi}}{dt^{-}_\lambda}=\sum\limits_{k=0}^\infty\Omega_{-k}\lambda^k,\quad \frac{d\bar{\phi}}{dt_{-k-2}}=\Omega_{-k},\qquad k\geq0,
\end{equation}
where
\begin{equation}\label{6.12}\begin{array}{lll}
\Omega_{-0}&=&\frac12 C_{N-1}\mathscr{A}_0,\quad \Omega_{-1}=\frac12(C_{N-2}\mathscr{A}_0+C_{N-1}\mathscr{A}_{-1}),\\
\Omega_{-2}&=&\frac12(C_{N-3}\mathscr{A}_0+C_{N-2}\mathscr{A}_{-1}+C_{N-1}\mathscr{A}_{-2}),\\
\Omega_{-k}&=&\frac12(C_{N-1-k}\mathscr{A}_0+C_{N-k}\mathscr{A}_{-1}+\cdots+C_{N-1}\mathscr{A}_{-k}),\quad 3\leq k\leq N-2,\\
\Omega_{-k}&=&\frac12(C_1\mathscr{A}_{-(k-N+2)}+C_2\mathscr{A}_{-(k-N+3)}+\cdots+C_{N-1}\mathscr{A}_{-k}),\quad k\geq N-1.\\
\end{array}
\end{equation}
\end{theorem}
{\bf Proof:} Recalling (\ref{5.9}), (\ref{6.7}), (\ref{4.21}), (\ref{4.22}) and (\ref{4.23}), by a direct calculation we arrive at
\begin{equation}\label{6.13}\begin{array}{ccl}
\displaystyle{\frac{d\bar{\phi}}{dt^{-}_\lambda}}&=&\displaystyle{-\frac12\frac{d\bar{\phi}}{d\tau^{-}_\lambda}=
-\frac12(C_1,C_2,\cdots,C_{N-1})
\left(\frac{d\tilde{\phi}_1}{d\tau_\lambda^{-}},
\frac{d\tilde{\phi}_2}{d\tau_\lambda^{-}},\cdots,\frac{d\tilde{\phi}_{N-1}}{d\tau_\lambda^{-}}\right)^T}\\
&=&\displaystyle{\frac{1}{2a(\lambda)}\sum\limits_{j=1}^{N-1}C_j\lambda^{N-1-j}}
=\displaystyle{\frac12\sum\limits_{k=0}^\infty\mathscr{A}_{-k}\lambda^k\sum\limits_{j=1}^{N-1}C_j\lambda^{N-1-j}}\\
&=&\displaystyle{\sum\limits_{k=0}^\infty\Omega_{-k}\lambda^k}.\\
\end{array}
\end{equation}
On the other hand, it is known from the expression of $(\ref{4.1})_2$ that
\begin{equation}\label{6.14}
\frac{d\bar{\phi}}{dt_\lambda^{-}}=\{\bar{\phi},H_\lambda^{-}\}_D
=\sum\limits_{k=0}^\infty\{\bar{\phi},H_{-k}\}_D\lambda^k=\sum\limits_{k=0}^\infty
\frac{d\bar{\phi}}{dt_{-k-2}}\lambda^k,
\end{equation}
which results in the formula $(\ref{6.11})_2$ by comparing the coefficients of $\lambda^k$ in (\ref{6.13}) and (\ref{6.14}).
\begin{theorem}
  \label{theo:4}
The forward Neumann flows are linearized by the Abel-Jacobi variable $\bar{\phi}$ on $J(\Gamma)$
\begin{equation}\label{6.15}
\frac{d\bar{\phi}}{dt^{+}_\lambda}=\sum\limits_{k=0}^\infty\Omega_{k}\lambda^{-k-1},\quad \frac{d\bar{\phi}}{dt_{k+1}}=\Omega_{k},\qquad k\geq0,
\end{equation}
where
\begin{equation}\begin{array}{lll}
\Omega_0&=&\frac{1}{2} C_1,\quad
\Omega_1=\frac{1}{2}(\bar{\Lambda}_1C_1+C_2),\quad \Omega_2=\frac12(\bar{\Lambda}_2C_1+\bar{\Lambda}_1C_2+C_3),\\
\Omega_k&=&\frac{1}{2}(\bar{\Lambda}_{k-1}C_1+\cdots+\bar{\Lambda}_1C_{k-1}+C_{k}),\quad 3\leq k\leq N-2,\\
\Omega_k&=&\frac{1}{2}(\bar{\Lambda}_{k-1}C_1+\cdots+\bar{\Lambda}_{k-N+2}C_{N-2}+\bar{\Lambda}_{k-N+1}C_{N-1}),\quad
k\geq N-1.
\end{array}\label{6.16}
\end{equation}
\end{theorem}

\par
As a concrete application of Theorems \ref{theo:3} and \ref{theo:4}, the Abel-Jacobi variable $\bar{\phi}$ is integrated by the direct quadrature on $J(\Gamma)$
\begin{equation}\label{6.17}
\bar{\phi}=\sum\limits_{k=0}^\infty\Omega_{-k}t_{-k-2}+\bar{\phi}_0+\sum\limits_{k=0}^\infty\Omega_{k}t_{k+1}.
\end{equation}
It has been shown in section \ref{sec:5} that each INLEEs staying in the bKdV hierarchy is reduced to two Neumann systems. By restricting $\bar{\phi}$ to finite number terms, the dynamic picture of various flows becomes clear, which signifies that the decomposition of quasi-periodic solutions of bKdV hierarchy can be reduced to linear superpositions along with flow variables
\begin{equation}
{\rm H_{-k}\ flow:}\quad \bar{\phi}=\bar{\phi}_0+\Omega_{-k}t_{-k-2},\qquad k\geq0,
 \label{6.18}\end{equation}
\begin{equation}
{\rm H_{k}\ flow:}\quad \bar{\phi}=\bar{\phi}_0+\Omega_{k-1}t_{k},\qquad k\geq1,
 \label{6.19}\end{equation}
\begin{equation}
{\rm X_{-k}\ flow:}\quad \bar{\phi}=\bar{\phi}_0+\Omega_0x+\Omega_{-(k-1)}t_{-k-1},\quad k\geq1,
 \label{6.20}\end{equation}
\begin{equation}
{\rm X_{k}\ flow:}\quad \bar{\phi}=\bar{\phi}_0+\Omega_0x+\Omega_kt_{k+1},\quad k\geq1.
 \label{6.21}\end{equation}
Due to the linearized flows (\ref{6.18})-(\ref{6.21}), we discuss the Riemann--Jacobi inversion from the Abel--Jacobi variable $\bar{\phi}$ to the elliptic variables $\{\mu_k\}_{k=1}^{N-1}$, which ultimately leads to Riemann theta function representations of the spectral potential $u$. It follows from (\ref{5.1}) and (\ref{5.2}) that the spectral potential $u$ can be described as the symmetric function of elliptic variables. We turn to the Riemann theorem \cite{42}, for the Abel--Jacobi variable $\bar{\phi}$ constructed by (\ref{6.7}), there exists a vector of Riemann constant $\mathcal {M}=(\mathcal {M}_1,\mathcal {M}_2,\cdots,\mathcal {M}_{N-1})^T$ such that $f(\lambda)=\theta({\cal{A}}(P(\lambda))-\bar{\phi}-\mathcal {M})$ has $N-1$ simple zeros at $\mu_1,\mu_2,\cdots,\mu_{N-1}.$ In order to make $f(\lambda)$ in the calculation well-defined, the Riemann surface $\Gamma$ should be properly cut along with the contours $a_j$ and $b_j$ to form a simply connected region with the boundary $\gamma$. With the help of the Residue theorem, the symmetric functions (negative and positive power sums) of $\{\mu_j\}_{j=1}^{N-1}$ can be attained by the inversion formulae
\begin{equation}\begin{array}{lll}
\displaystyle{\sum\limits_{j=1}^{N-1}\mu_j^{-k}}&=&\displaystyle{I_{-k}(\Gamma)-
\sum\limits_{s=1}^2\underset{\lambda=0_s}{\rm Res}\lambda^{-k}
d\ln f(\lambda)},\vspace{1mm}\\
\displaystyle{\sum\limits_{j=1}^{N-1}\mu_j^k}&=&\displaystyle{I_k(\Gamma)-
\underset{\lambda=\infty}{\rm Res}\lambda^k d\ln f(\lambda)},\\
\end{array}
\label{6.22}\end{equation}
where $I_{-k}(\Gamma)=\sum\limits_{j=1}^{N-1}\int_{a_j}\lambda^{-k}
\omega_j$ and $I_k(\Gamma)=\sum\limits_{j=1}^{N-1}\int_{a_j}\lambda^k
\omega_j$ are two constants independent of $\bar{\phi}$ \cite{43}.

\par
In the neighbourhood of $\lambda=\infty$, it follows from the local coordinate $\lambda=z^{-2}$ that the normalized basis of holomorphic differential $\omega$ can be expanded as
\begin{equation}\label{6.23}
\begin{array}{lll}
\omega&=&(C_1,C_2,\cdots,C_{N-1})(\tilde{\omega}_1,\tilde{\omega}_2,\cdots,\tilde{\omega}_{N-1})^T\\
&=&\frac{\lambda^{N-1}}{4\sqrt{-R(\lambda)}}\sum\limits_{j=1}^{N-1}C_j\lambda^{-j}d\lambda\\
&=&\frac{\lambda^{-\frac12}}{4\sqrt{-1}}\sum\limits_{k=0}^\infty\bar{\Lambda}_k\lambda^{-k}\sum\limits_{j=1}^N C_j\lambda^{-j}d\lambda\\
&=&\sqrt{-1}\sum\limits_{k=0}^\infty\Omega_kz^{2k}dz,
\end{array}
\end{equation}
which gives rise to the asymptotic expansion of ${A}(P(\lambda))$ at $z=0$
\begin{equation}
\mathcal {A}(P(\lambda))=-\chi
+\sqrt{-1}\sum\limits_{k=0}^\infty \frac{\Omega_kz^{2k+1}}{2k+1} ,\qquad \chi=\int_{\infty}^{P_0}\omega.
\label{6.24}\end{equation}
Let $\varsigma_j$ be the $j$th component of $f(\lambda)$, $\partial_j=\partial/\partial\varsigma_j$, $\partial^2_{jk}=\partial^2/\partial\varsigma_j\partial\varsigma_k$, etc. With the Einstein summation convention, in the local coordinate $f(\lambda)$ has the Maclaurin expansion
\begin{equation}\begin{array}{l}
f(\lambda)=f(z^{-2})=\theta^{(\infty)}(\bar{\phi}+M+\chi)
-\sqrt{-1}\Omega_{0j}\partial_j\theta^{(\infty)}z\\
-\frac12\Omega_{0j}\Omega_{0k}\partial^2_{jk}\theta^{(\infty)}z^2-
\frac{\sqrt{-1}}{6}(2\Omega_{1j}\partial_{j}\theta^{(\infty)}-\Omega_{0j}\Omega_{0k}\Omega_{0l}
\partial^3_{jkl}\theta^{(\infty)})z^3\\
+\frac{1}{24}(\Omega_{0j}\Omega_{0k}\Omega_{0l}\Omega_{0m}\partial^4_{jklm}\theta^{(\infty)}
-8\Omega_{1j}\Omega_{0k}\partial^2_{jk}\theta^{(\infty)})z^4+o(z^4),\\
\end{array}\label{6.25}
\end{equation}
which indicates that
\begin{equation}\begin{array}{lll}
\frac{d\ln f(z^{-2})}{dz}&=&\frac{\Omega_{0j}\partial_j\theta^{(\infty)}}{\sqrt{-1}\theta^{(\infty)}}+
\left(\left(\frac{\Omega_{0j}\partial_j\theta^{(\infty)}}{\theta^{(\infty)}}\right)^2-\frac{1}{\theta^{(\infty)}}
\Omega_{0j}\Omega_{0k}\partial^2_{jk}\theta^{(\infty)}\right)z\\
&&-\frac{\sqrt{-1}}{2}\left(\frac{2\Omega_{1j}\partial_j\theta^{(\infty)}}{\theta^{(\infty)}}+\frac{3}
{(\theta^{(\infty)})^3}\Omega_{0j}\partial_j\theta^{(\infty)}\Omega_{0k}\Omega_{0l}\partial^2_{kl}\theta^{(\infty)}
\right.\\&&\left.-2\left(\frac{\Omega_{0j}\partial_j\theta^{(\infty)}}{\theta^{(\infty)}}\right)^3
-\frac{1}{\theta^{(\infty)}}\Omega_{0j}\Omega_{0k}\Omega_{0l}\partial^3_{jkl}\theta^{(\infty)}\right)z^2
\\&&+\frac{1}{6}\left(\frac{1}{\theta^{(\infty)}}\Omega_{0j}\Omega_{0k}\Omega_{0l}\Omega_{0m}\partial^4_{jklm}\theta^{(\infty)}
-\frac{8}{\theta^{(\infty)}}\Omega_{0j}\Omega_{1k}\partial^2_{jk}\theta^{(\infty)}\right.
\\&&-\frac{3}{(\theta^{(\infty)})^2}(\Omega_{0j}\Omega_{0k}\partial^2_{jk}\theta^{(\infty)})^2+
\frac{8}{(\theta^{(\infty)})^2}\Omega_{0j}\partial_j\theta^{(\infty)}\Omega_{1k}\partial_k\theta^{(\infty)}
\\&&+\frac{12}{\theta^3}(\Omega_{0j}\partial_j\theta^{(\infty)})^2\Omega_{0k}\Omega_{0l}\partial^2_{kl}\theta^{(\infty)}-
6\left(\frac{\Omega_{0j}\partial_j\theta^{(\infty)}}{\theta^{(\infty)}}\right)^4
\\&&\left.-\frac{4}{(\theta^{(\infty)})^2}\Omega_{0j}\partial_j\theta^{(\infty)}
\Omega_{0k}\Omega_{0l}\Omega_{0m}\partial^3_{klm}\theta^{(\infty)}\right)z^3+o(z^3).\\
\end{array}\label{6.26}
\end{equation}
And thus, by (\ref{6.22}) and (\ref{6.26}), we come to the trace formulae
\begin{equation}\begin{array}{c}
\sum\limits_{k=1}^{N-1}\mu_k=I_1(\Gamma)+\partial^2\ln\theta(\bar{\phi}+M+\chi),\\
\sum\limits_{k=1}^{N-1}\mu^2_k=I_2(\Gamma)+\frac43\partial_{xt_2}^2\ln\theta(\bar{\phi}+M+\chi)-\frac14
\partial^4\ln\theta(\bar{\phi}+M+\chi),\\
\cdots\cdots,\ \rm{etc}.\\
\end{array}\label{6.27}
\end{equation}

\par
Only for the succinctness in writing, let us make the notation
\begin{equation}\label{6.28}
\alpha=2I_1(\Gamma)-2\sum\limits_{j=1}^N\lambda_j+F_1,\qquad \kappa=\bar{\phi}_0+M+\chi.
\end{equation}
In the end, resorting to Proposition \ref{prop:2}, Theorems \ref{theo:3}-\ref{theo:4}, (\ref{5.1}), (\ref{5.2}), (\ref{6.27}), as well as Theorem 2 in \cite{36}, we obtain
\begin{itemize}
\item the quasi-periodic solutions of nKdV hierarchy (\ref{2.23})
\begin{equation}\label{6.29}
u(x,t_{-k-2})=\alpha+2\partial^2\ln\theta(\Omega_0x+\Omega_{-k}t_{-k-2}+\kappa),\qquad k\geq0,
\end{equation}
where the case of $k=0$ is the quasi-periodic solution of nKdV equation (\ref{1.1}) (or (\ref{1.2}));
\item and the quasi-periodic solutions of pKdV hierarchy (\ref{2.15})
\begin{equation}\label{6.30}
u(x,t_{k+1})=\alpha+2\partial^2\ln\theta(\Omega_0x+\Omega_{k}t_{k+1}+\kappa),\qquad k\geq1,
\end{equation}
where the case of $k=1$ is the quasi-periodic solution of KdV equation (\ref{2.17}).
\end{itemize}

\setcounter{equation}{0}
\begin{appendices}
\section{Quasi-periodic solutions to the Kupershmidt deformation of KdV hierarchy}
\label{sec:A}
Note that the KdV6 equation and its generalization are indeed the integrable couplings of pKdV and nKdV flows. We point out that they can be solved incidentally by using the backward and forward Neumann systems. 

\par
According to \cite{5}, the so-called KdV6 equation takes the form
\begin{equation}\label{7.1}
(\partial^3+4w_x\partial+2w_{xx})(w_{t_{(2,-2)}}+\frac14(w_{xxx}+6w_x^2))=0,
\end{equation}
which is equivalent to
\begin{equation}\label{7.2}
\left\{\begin{array}{l}
u_{t_{(2,-2)}}+\frac14(u_{xxx}+6uu_x)+2v_x=0,\\
v_{xxx}+4uv_x+2u_xv=0,\\
\end{array}
\right.
\end{equation}
under the transformation
\begin{equation}\label{7.3}
u=w_x,\qquad v=-\frac12w_{t_{(2,-2)}}-\frac18(w_{xxx}+6w_x^2).
\end{equation}
Resorting to two Hamiltonian operators (or the Lenard operator pair $K$ and $J$), Kupershmidt described the KdV6 equation (\ref{7.1}) as a nonholonomic deformation of bi-Hamiltonian system \cite{6}
\begin{equation}\label{7.4}
u_{t_{(2,-2)}}=J\frac{\delta\mathscr{H}_1}{\delta u}-J(v)=K\frac{\delta\mathscr{H}_2}{\delta u}-J(v),
\end{equation}
\begin{equation}\label{7.5}
\mathscr{H}_1=\frac{1}{16}u_x^2-\frac18u^3,\qquad \mathscr{H}_2=\frac14u^2.
\end{equation}
Based on the construction of nonholonomic deformation, Zhou further generalized it to the mixed hierarchy of soliton equations \cite{7}
\begin{equation}\label{7.6}
\left\{\begin{array}{l}
u_{t_{(n,-m)}}=\mathcal{C}_1Jg_{n-1}+\mathcal{C}_2Jg_{-m-1},\qquad n\geq2,\ m\geq2,\\
Kg_k=Jg_{k+1},\qquad k\leq-4,\ \rm{or}\ k\geq-1,\ k\in\mathbb{Z},
\end{array}
\right.
\end{equation}
which includes the Kupershmidt deformation as its special member.

\par
It follows from \cite{7} that the KdV6 equation (\ref{7.1}) has the Lax representations (\ref{2.1}) and
\begin{equation}
\label{7.7}
\varphi_{t_{(2,-2)}}=V^{(2,-2)}\varphi,\qquad V^{(2,-2)}=V^{(2)}+V^{(-2)},
\end{equation}
and the Kupershmidt deformation of KdV hierarchy (\ref{7.6}) has the Lax representations (\ref{2.1}) and
\begin{equation}
\label{7.8}
\varphi_{t_{(n,-m)}}=V^{(n,-m)}\varphi,\qquad V^{(n,-m)}=\mathcal{C}_1V^{(n)}+\mathcal{C}_2V^{(-m)}.
\end{equation}
Followed by the Neumann map (\ref{3.9}), the KdV6 equation is reduced to the Neumann system (\ref{3.13}) and the mixed Neumann system
\begin{equation}\label{7.9}
p_{t_{(2,-2)}}=\{p,H_{(2,-0)}\}_D,\quad q_{t_{(2,-2)}}=\{q,H_{(2,-0)}\}_D,\quad H_{(2,-0)}=H_2+H_{-0}.
\end{equation}
The Kupershmidt deformation of KdV hierarchy is reduced to the Neumann system (\ref{3.13}) and the mixed Neumann system
\begin{equation}\label{7.10}
p_{t_{(n,-m)}}=\{p,H_{(n,-m+2)}\}_D,\qquad q_{t_{(n,-m)}}=\{q,H_{(n,-m+2)}\}_D,
\end{equation}
where
\begin{equation}\label{7.11}
H_{(n,-m+2)}=\mathcal{C}_1H_n+\mathcal{C}_2H_{-m+2}.
\end{equation}

Recalling Theorems \ref{theo:3} and \ref{theo:4}, the Abel-Jacobi variable $\bar{\phi}$ straightens out the KdV6 flow
\begin{equation}
\label{7.12}
\bar{\phi}=\bar{\phi}_0+\Omega_0x+(\Omega_{-0}+\Omega_1)t_{(2,-2)},
\end{equation}
which results in its quasi-periodic solution
\begin{equation}\label{7.13}
w(x,t_{(2,-2)})=\alpha x+2\partial\ln\theta(\Omega_0x+(\Omega_{-0}+\Omega_1)t_{(2,-2)}+\kappa),
\end{equation}
and the Abel-Jacobi variable $\bar{\phi}$ straightens out the mixed KdV flow
\begin{equation}
\label{7.14}
\bar{\phi}=\bar{\phi}_0+\Omega_0x+(\mathcal{C}_1\Omega_{-m+2}+\mathcal{C}_2\Omega_{n-1})t_{(n,-m)},\quad n,\ m\geq3,
\end{equation}
which yields its quasi-periodic solution
\begin{equation}\label{7.15}
u(x,t_{(n,-m)})=\alpha+2\partial^2\ln\theta(\Omega_0x+
(\mathcal{C}_1\Omega_{-m+2}+\mathcal{C}_2\Omega_{n-1})t_{(n,-m)}+\kappa),\quad n,\ m\geq3.
\end{equation}
\end{appendices}

\section*{Acknowledgements}
This work was supported by the National Natural Science Foundation
of China (No.11471072).

\end{document}